\documentclass{article}

\usepackage{mathrsfs}
\usepackage{amsfonts}
\usepackage{amsmath}
\usepackage{amssymb}
\usepackage{graphicx}

\title{Mathematics and explanation in astronomy and astrophysics}
\author{Gordon McCabe}

\def\eqalign#1{\,\vcenter{\openup.7ex\mathsurround=0pt
 \ialign{\strut\hfil$\displaystyle{##}$&$\displaystyle{{}##}$\hfil
 \crcr#1\crcr}}\,}

\begin{document}

\maketitle

\begin{abstract}

The purpose of this paper is to expound and clarify the
mathematics and explanations commonly employed in certain notable
areas of astronomy and astrophysics. The first section
concentrates upon the mathematics employed to represent and
understand stellar structure and evolution. The second section
analyses two different explanations for the structure of spiral
galaxies.

\end{abstract}

\section{Stellar structure and evolution}

A star forms when a body of gas contracts under its own
gravitational attraction, and the pressure and temperature created
at the centre of the agglomerated mass is sufficient to ignite the
nuclear fusion of atomic nuclei. The radiation released as a
by-product of the nuclear fusion eventually reaches the surface of
the star and becomes starlight.

Newtonian astrophysics represents the gaseous content of a star as
a fluid which occupies a compact open subset $\Omega \subset
\mathbb{R}^3$ of three-dimensional Euclidean space. The fluid is
considered to possess a mass density scalar field $\rho$ and a
pressure scalar field $p$. To study stellar structure in Newtonian
astrophysics, it is common to assume that the star is spherically
symmetric and static (time-independent). The requirement that the
star be static is equivalent to the requirement that the star is
in hydrostatic and thermal equilibrium. For a static spherically
symmetric star, the fluid occupies a solid ball of radius $r$, the
mass density and pressure are time independent, and the velocity
vector of the fluid vanishes. Using spherical polar coordinates,
$(r,\theta,\phi)$, the pressure and the density are independent of
the angular coordinates $(\theta,\phi)$, but vary as a function of
the radius $r$. The fluid is assumed to be self-gravitating and to
satisfy the Poisson equation of Newtonian gravitation,

$$\nabla^2 \Phi = - 4 \pi G \rho \;,$$ with respect to a
gravitational potential scalar field $\Phi$. $\nabla^2$ is the
Laplacian, also denoted as $\Delta$ in many texts.

There are four differential equations which govern the structure
of a static, spherically symmetric star. The four equations taken
together constitute a coupled (`simultaneous') set of ordinary
first-order, \emph{non-linear} differential equations. We now
proceed to introduce and discuss these equations.

Assuming spherical symmetry and time-independence, the mass
density is a function of radius alone $\rho(r)$, and from the
definition of mass density and Euclidean geometry, it follows that
$m(r)$, the mass enclosed within the surface of radius $r$, is
given by the expression:

$$
m(r) = \int_0^r 4 \pi (r')^2 \rho(r') dr' \;.
$$ (The prime here simply indicates the use of a dummy variable for the integration.)
This gives us our first differential equation for stellar
structure, sometimes called the mass continuity equation:

$$
\frac{dm(r)}{dr} = 4 \pi r^2 \rho(r) \;.
$$

The Newtonian gravitational potential for a self-gravitating,
spherically symmetric body satisfying the Poisson equation, is
$\Phi = -Gm(r)r^{-1}$. Given a gravitational potential $\Phi$, the
gravitational force exerted per unit mass is specified by the
vector field $ - \text{grad} \; \Phi$, and the gravitational force
per unit volume is specified by $F = - \rho \; \text{grad} \;
\Phi$. Most texts denote the gradient operator as $\nabla$, but
this can be confused with the covariant derivative operator in
differential geometry. Whilst the gradient of a scalar field is a
contravariant vector field, the covariant derivative of a scalar
field is a \emph{covariant} vector field. Hence, to avoid
confusion, we shall write `$\text{grad}$' rather than $\nabla$.
Given a choice of coordinates $(x_1,...,x_n)$ on a manifold, and
given a metric tensor field $g_{ij} dx^i \otimes dx^j$ which
specifies the geometry of space, the gradient of an arbitrary
scalar field $f$ can be expressed as

$$
\text{grad} \; f = g^{ij} \frac{\partial f}{\partial
x_i}\frac{\partial}{\partial x_j} \;,
$$ where $g^{ij}$ is the inverse of the matrix $g_{ij}$. In the
spherical polar coordinates we have chosen, $(x_1,x_2,x_3) =
(r,\theta,\phi)$, the flat Euclidean metric takes the form

$$
ds^2 = dr \otimes dr + r^2 d\theta \otimes d \theta +
r^2\sin^2\theta d\phi \otimes d\phi \;,
$$ hence the gradient of $f$ takes the form

$$
\text{grad} \; f = \frac{\partial f}{\partial
r}\frac{\partial}{\partial r} + r^2 \frac{\partial f}{\partial
\theta}\frac{\partial}{\partial \theta} + r^2 \sin^2 \theta
\frac{\partial f}{\partial \phi}\frac{\partial}{\partial \phi} \;.
$$ Assuming spherical symmetry of the gravitational potential
$\Phi$ entails that $\partial \Phi/\partial \theta$ and $\partial
\Phi/\partial \phi$ vanish, hence

$$\eqalign{\text{grad} \; \Phi &= \frac{d \Phi}{dr} \frac{\partial}{\partial r}
\cr &= \frac{d (-Gm(r)r^{-1})}{dr} \frac{\partial}{\partial r} \cr
&= Gm(r)r^{-2} \frac{\partial}{\partial r} \;.}$$

The requirement of hydrostatic equilibrium corresponds to the
requirement that the inward gravitational force $ - \rho \;
\text{grad} \; \Phi$ is balanced at each point by the outward
pressure gradient force $ - \text{grad} \; p$:\footnote{The
pressure at each radius is a sum of the gas pressure and the
radiation pressure.}

$$
\text{grad} \; p + \rho \: \text{grad} \; \Phi = 0 \;.
$$ Spherical symmetry entails that $\partial p/\partial \theta$ and
$\partial p/\partial \phi$ vanish, hence

$$
\text{grad} \; p =  \frac{dp}{dr}\frac{\partial}{\partial r} =
-Gm(r)\rho(r) r^{-2}\frac{\partial}{\partial r} \;.
$$ This gives us our second differential equation for stellar
structure:

$$
\frac{dP(r)}{dr} = - \frac{Gm(r)\rho(r)}{r^2} \;.
$$

Let $l(r)$ denote the amount of energy passing through the surface
of radius $r$ per unit time. Let $\epsilon(r)$ denote the energy
production coefficient, the amount of energy released at radius
$r$ in the star per unit time and per unit mass. It follows that
$4 \pi r^2 \rho(r) \epsilon (r)$ is the energy released per unit
time at radius $r$. A star in thermal equilibrium is an open
system for which there are large energy flows out of the system,
and in which there are significant temperature and pressure
gradients, but in which, nevertheless, the temperature and
pressure profile of the star remain constant. For a star in
thermal equilibrium, none of the energy released is used to heat
up the star or change its volume. Hence, for such a star, the rate
of energy flow at radius $r$ is given by the expression:

$$
l(r) = \int_0^r 4 \pi (r')^2 \rho(r') \epsilon(r') dr' \;.
$$ This gives us our third differential equation of stellar
structure,

$$
\frac{dl(r)}{dr} = 4 \pi r^2 \rho(r) \epsilon(r) \;.
$$

For a star of radius $R$ the \emph{luminosity} is defined to be $L
= l(R)$, the amount of energy passing through the outer surface of
the star per unit time. When a star resides in a state of thermal
equilibrium, the total amount of energy produced by the star per
unit time equals the amount of energy radiated from the outer
surface per unit time. Hence, the luminosity $L$ of such a star is
given by the equation:

$$
L = \int_0^R 4 \pi r^2 \rho(r) \epsilon(r) dr \;.
$$

The energy per unit time $l(r)$ passing through the sphere of
radius $r$ can be expressed as

$$
l(r) = 4 \pi r^2 F(r) \;,
$$ where $F$ is the energy flux, the energy flow per unit time per unit
area. When there is a temperature gradient in a body, energy flow
will occur by conduction and radiation at the very least.
Assuming, for simplicity, the absence of convection, the energy
flux in a star can be broken down into a conductive energy flux
$F_{cond}(r)$ and a radiative flux $F_{rad}(r)$. In both cases,
the energy flow is proportional to the temperature gradient as
follows (Tayler 1994, p63):

$$
F_{cond}(r) = -
\frac{4acT^3(r)}{3\kappa_{cond}(r)\rho(r)}\frac{dT(r)}{dr} \;,
$$ and

$$
F_{rad}(r) = -
\frac{4acT^3(r)}{3\kappa_{rad}(r)\rho(r)}\frac{dT(r)}{dr} \;.
$$ $\kappa_{cond}$, the coefficient of conductive opacity, measures the resistance to the flow of heat by conduction, and
$\kappa_{rad}$, the coefficient of radiative opacity, measures the
resistance to the flow of heat by radiation. $a$ is the radiation
density constant, $a= 7.57 \times 10^{-15} \; \text{erg cm}^{-3}
\text{K}^{-4}$.

Assuming that energy flow in a star is due to conduction and
radiation,\footnote{Conduction is only significant in `compact'
stars, i.e., white dwarf and neutron stars.} one obtains (Tayler
1994, p64):

$$
l(r) = 4\pi r^2(F_{cond}(r) + F_{rad}(r)) = - \frac{16\pi
acr^2T^3(r)}{3\kappa(r) \rho(r)}\frac{dT(r)}{dr} \;,
$$ where

$$
\frac{1}{\kappa(r)} = \frac{1}{\kappa_{cond}(r)} +
\frac{1}{\kappa_{rad}(r)} \;.
$$ Re-arranging, one obtains the following differential equation
for the temperature gradient in a star, which is the fourth
equation governing stellar structure:

$$
\frac{dT(r)}{dr} = - \frac{3\kappa(r) \rho(r)}{4 ac T^3(r)}
\frac{l(r)}{4 \pi r^2} \;.
$$

The chemical composition of a spherically symmetric and static
star can be specified by a set of functions $X_i(r)$, which
specify the fraction of unit mass consisting of nuclei of type
$i$, for $i = 1,...,I$. As such, $\Sigma_i X_i(r) =
1$.\footnote{It is a common approximation to use only three
fractions $(X,Y,Z)$, which are such that $X+Y+Z =1$, and which
specify, respectively, the proportion of hydrogen, the proportion
of helium, and the proportion of all other elements in the
constitution of a star.}

Before the equations of stellar structure can be solved, one must
specify an equation of state for the pressure,

$$
P = P(\rho, T, X_1,...,X_I) \;,
$$ an expression for the energy production coefficient

$$
\epsilon = \epsilon(\rho, T, X_1,...,X_I) \;,
$$ and an expression for the opacity

$$
\kappa = \kappa(\rho, T, X_1,...,X_I) \;.
$$ These expressions are sometimes called the constitutive
equations.

If these three functions are specified, and four boundary
conditions are specified, then one can solve the four differential
equations of stellar structure. The four boundary conditions can
be chosen to be:

$$
m =0 \;, \quad  l= 0 \quad \text{at} \;\; r =0 \;,
$$ and

$$
\rho = 0 \;, \quad T = 0 \quad \text{at} \;\; r = R \;.
$$ Note here that the star is idealised as an \emph{open} solid
ball of radius $R$, hence the boundary of the star at radius $R$
is not itself considered to be part of the star. Choosing $\rho =
0$ (or $P=0$) and $T = 0$ at a finite radius $r = R$ is merely a
simple idealisation. In reality, the density, pressure and
temperature of a star gradually descend to the non-zero value of
the interstellar medium in the neighbourhood of the star.

Solving the four differential equations for stellar structure then
gives the mass, pressure, energy flow and temperature as a
function of radius. However, in practical terms, one might wish to
specify the total mass $M$ and chemical composition of a star, and
then obtain, amongst other things, the radius of such a star in
hydrostatic and thermal equilibrium. It is therefore conventional
to re-cast the differential equations with mass $m$, rather than
radius $r$, as the independent variable. Doing so, one obtains
(Tayler 1994, p70-71):

$$
\frac{dr(m)}{dm} = \frac{1}{4 \pi r^2(m) \rho(m)} \;,
$$

$$
\frac{dP(m)}{dm} = - \frac{Gm}{4 \pi r^4(m)} \;,
$$

$$
\frac{dl(m)}{dm} = \epsilon(m) \;,
$$

$$
\frac{dT(m)}{dm} = - \frac{3\kappa(m) l(m)}{64 \pi^2 ac r^4(m)
T^3(m)} \;.
$$ The chemical composition of the star is then expressed in terms
of functions $\{X_i(m): i=1,...,I \}$.

One can then choose boundary conditions for the re-formulated
equations of stellar structure, such as:

$$
r =0 \;, \quad  l= 0 \quad \text{at} \;\; m =0 \;,
$$ and

$$
\rho = 0 \;, \quad T = 0 \quad \text{at} \;\; m = M \;.
$$ Again, one could choose $P =0$ at $m=M$ rather than $\rho = 0$,
and, again, this is merely the simplest idealisation on offer.

Now consider the relationship between theory and observation. The
observational state of a star at a moment in time can be specified
by just two parameters: its luminosity and `surface' temperature.
Let us take these in turn:

\begin{enumerate}
\item{If one calculates the distance to an
observable star, say by parallax, then one can infer the
luminosity of the star from its apparent luminosity.}

\item{Whilst the temperature of the surface of a star was idealised
in the boundary conditions above to be zero, a more sophisticated
model divides a star into its interior and atmosphere. The lowest
level of the atmosphere is called the photosphere, and the
`surface' temperature of a star is deemed to be the temperature of
the photosphere. The temperature of a star's surface, in this
sense, is intimately related to the spectral type of the star,
defined by the absorption lines in the spectrum of light emitted
by the photosphere. Given that the temperature of a star's
photosphere largely determines the type of its spectrum, the
observational state of a star can be specified by its luminosity
and spectral type. There are ten spectral types which, in order of
decreasing temperature, are referred to as O, B, A, F, G, K, M, R,
N, and S.} \end{enumerate}

If one knows both the luminosity and temperature of a star, then
one can calculate the radius of the star. The luminosity of a star
is a function of the radius $R$ of the star and the
effective\footnote{Strictly, the effective surface temperature
$T_e$ is defined as the temperature of a black body with the same
radius and luminosity as the star.} surface temperature $T_e$ of
the star, according to the following expression,

$$
L = 4 \pi R^2 \sigma T_e^4 \;,
$$ where $\sigma$ is the Stefan-Boltzmann constant, related to the radiation density constant
$a$ by $\sigma = ac/4$. Given $L$ and $T_e$, one can obviously use
this expression to calculate $R$.

The observational state of a star can be represented by a point on
the Hertzsprung-Russell diagram, a two-dimensional rectangle,
coordinatized by luminosity on the vertical axis, and either
spectral type or temperature on the horizontal axis. The
observational history of a star traces out a path on the
Hertzsprung-Russell diagram.

The luminosity and temperature of a star in hydrostatic and
thermal equilibrium are the observational properties which can be
explained by solving the differential equations of stellar
structure. This explanation falls under the aegis of the
deductive-nomological (D-N) account of scientific explanation. In
such explanations, one explains certain phenomenal facts by
logically deriving them from the conjunction of general laws and
particular specified circumstances. In this case, the mass and
chemical composition of a star are the specified circumstances,
and given a specification of the constitutive equations, one can
explain the luminosity and temperature of a star from the
conjunction of the equations of stellar structure and the given
mass and chemical composition.

The initial mass and chemical composition of different stars vary,
and these are the characteristics which determine the history and
lifetime of a star on the Hertzsprung-Russell diagram. Stars which
are burning hydrogen into helium, in a state of hydrostatic and
thermal equilibrium, occupy a roughly diagonal channel, running
from the top left to the bottom right of the Hertzsprung-Russell
diagram, called the \emph{main sequence}. The initial mass, and to
a lesser extent, the initial chemical composition of a star,
determine its initial location in the main sequence. Stars with
greater initial mass are more luminous, and occupy positions
further up the main sequence slope. The larger the mass and
luminosity of a star, the shorter the length of time it will spend
on the main sequence until its hydrogen fuel is expended.

The Vogt-Russell `theorem', conjectured independently by Heinrich
Vogt (1926) and Henry Norris Russell (Russell \textit{et al} 1927,
p910), states that with the total mass and chemical composition
specified, the equations of stellar structure admit a unique
solution. If true, the Vogt-Russell theorem would entail that each
mass and chemical composition corresponds to a unique
\emph{equilibrium} configuration. This, of course, is consistent
with the fact that stars of the same mass but different chemical
profile, can possess different equilibrium states. The chemical
composition of a star will change during its lifetime as nuclear
fusion converts lighter elements into heavier elements, and a star
will tend to pass through a sequence of different equilibrium
configurations.

However, whilst the Vogt-Russell theorem is essentially true, it
is not strictly true, and it is certainly no theorem. A system of
four coupled, \emph{non-linear}, ordinary differential equations,
with boundary conditions specified at two different points, does
not necessarily admit a unique solution. Whilst there is usually a
unique equilibrium configuration for a given mass and chemical
composition, for some combinations of mass and chemical
composition there are multiple solutions to the equations of
stellar structure. The notion that there was actually a `theorem'
was promulgated in textbooks, and never given a rigorous proof. As
a conjecture, it is now known to be false. In fact, both the
existence and uniqueness parts of the conjecture fail. For
example, there is no solution composed of helium with a mass less
than about 0.3 solar masses, and Cox and Salpeter (1964)
demonstrated that there are two different solutions for stars of
the same mass, burning purely helium, close to this minimum mass.
This is referred to as the \emph{double-valued} helium main
sequence, in the sense that, for some range of masses, two values
of radius $R$ are possible for each value of mass. K\"ahler (1978)
tracked the post main sequence evolution of the helium core of a
star of two solar masses, and found different equilibrium
configurations of different radius, but with the same mass.
Gabriel and Noels-Gr\"otsch (1968) demonstrated that the minimum
mass of a solution composed entirely of carbon is about 0.9 solar
masses, and that, once again, there are two possible solutions for
each mass value greater than the minimum mass. In one branch of
such double-valued main sequences, the radius and luminosity
increase with increasing mass, whilst in the other branch, there
are smaller radii, and the luminosity decreases with increasing
mass (Hansen 1978, p23). The latter, anomalous branches,
correspond physically to the presence of electron degeneracy. It
is therefore not strictly true to say that the mass and chemical
composition of a star in equilibrium uniquely determine its
structure. However, as Hansen \textit{et al} point out, ``the idea
of uniqueness is still useful in that among a set of models all
having the same mass and run of composition, usually only one
seems to correspond to a real star or to have come from some
realistic line of stellar evolution. The others are unstable in
some fundamental way (as far as we know)," (2004, p331).

Given a fixed mass and chemical composition, each solution to the
differential equations of stellar structure corresponds to a pair
of values $(P_c,T_c)=(P(0),T(0))$ for the central pressure and
temperature of the star, or, equivalently, to a pair of values
$(R,L)=(r(M),L(M))$ for the surface radius and luminosity. The
significance of this is that the radius $R$ and luminosity $L$ of
a star are both inferrable from observation. If the mass and
chemical composition determined a unique solution, then the mass
and chemical composition of a star in equilibrium would uniquely
determine its radius and luminosity $(R,L)$. If, on the contrary,
there is no unique solution for a particular combination of mass
and chemical composition, then there are multiple corresponding
pairs $\{(R,L)_i: i =1,...,n\}$.

If one drops the requirements of hydrostatic and thermal
equilibrium, then one can obtain the equations of stellar
structure and evolution which govern the history of a Newtonian
star. Assuming spherical symmetry, the equations are (Kippenhahn
and Weigert 1990, p64):

$$
\frac{\partial r(m,t)}{\partial m} = \frac{1}{4 \pi r^2(m,t)
\rho(m,t)} \;,
$$

$$
\frac{\partial P(m,t)}{\partial m} = - \frac{Gm}{4 \pi r^4(m,t)} -
\frac{1}{4 \pi r^2(m,t)}\frac{\partial^2 r(m,t)}{\partial t^2} \;,
$$

$$
\frac{\partial l(m,t)}{\partial m} = \epsilon(m) - c_P
\frac{\partial T(m,t)}{\partial t} + \frac{\delta(m,t)}{\rho(m,t)}
\frac{\partial P(m,t)}{\partial t}\;,
$$

$$
\frac{\partial T(m,t)}{\partial m} = - \frac{3\kappa(m,t)
l(m,t)}{64 \pi^2 ac r^4(m,t) T^3(m,t)} \;.
$$ Notice that these are \emph{partial} differential equations,
whilst the equations of stellar structure are \emph{ordinary}
differential equations. $c_P$ is the specific heat at constant
pressure and $\delta \equiv -(\partial \ln \rho/\partial \ln
T)_P$, (see Kippenhahn and Weigert, p19). The fourth equation here
continues to assume that energy transport is due to radiation and
conduction alone, but this equation can be generalized.

Time-dependence requires one to introduce additional equations for
the time-dependence of the mass fractions $X_i$:

$$
\frac{\partial X_i(m,t)}{\partial t} = \frac{m_i}{\rho(m,t)}\left
(\Sigma_j r_{ji} - \Sigma_k r_{ik} \right ) \;, \quad \quad i =
1,...,I \;.
$$ $m_i$ is the mass of the nuclei of type $i$, and $r_{ij}$ is
the rate at which nuclei of type $i$ are transformed into nuclei
of type $j$ per unit volume.

Solving these equations of stellar structure and evolution
requires the specification of initial conditions as well as
boundary conditions.

The histories of different star types can intersect when plotted
as paths on the Hertzsprung-Russell diagram. For example, after a
star of around one solar mass has expended all the hydrogen in its
core, it leaves the main sequence, increasing in radius and
decreasing in temperature, to the effect of a net increase in
luminosity. This evolutionary path takes the star up the
Hertzsprung-Russell diagram to become a red giant. It then
subsequently increases in temperature, crossing the main sequence
itself from right to left, before it sheds its outer layers in the
form of a so-called planetary nebula. Where this path crosses the
main sequence, the mass of the star is considerably below the mass
of stars at that point on the main sequence. If two different
types of star can occupy the same point on the Hertzsprung-Russell
diagram, it entails that knowledge of the luminosity and
temperature of a star alone is not sufficient to determine the
unique evolutionary history of a star. The Hertzsprung-Russell
diagram does not provide a state space for stars in the sense of a
state space (`phase space') in Hamiltonian mechanics.

Nuclear fusion inside stars is responsible for creating almost all
of the atoms in our universe which are heavier than hydrogen or
helium. These heavier elements are generically called `metals' by
astrophysicists. During their lifetime, and particularly at the
end of their lifetime, stars will eject a proportion of their mass
into the interstellar medium, and this mass will contain, in some
proportion, the metallic elements created by fusion. The mass
returned into the interstellar medium is re-cycled in the
formation of subsequent generations of stars.

The stars in our galaxy are classified, in a coarse-grained
fashion, as either Population $\rm{I}$ or Population $\rm{II}$.
The Population $\rm{II}$ stars were the first generation of stars
formed in our galaxy, and as such, are metal-poor stars.
Population $\rm{I}$ stars were formed from an interstellar medium
which already contained the metallic elements created by the first
generation of stars, hence the Population $\rm{I}$ stars have a
greater proportion of metallic elements in their chemical
constitution.

\section{Galaxies}

\begin{figure}[h]
\centering
\includegraphics[scale = 0.5]{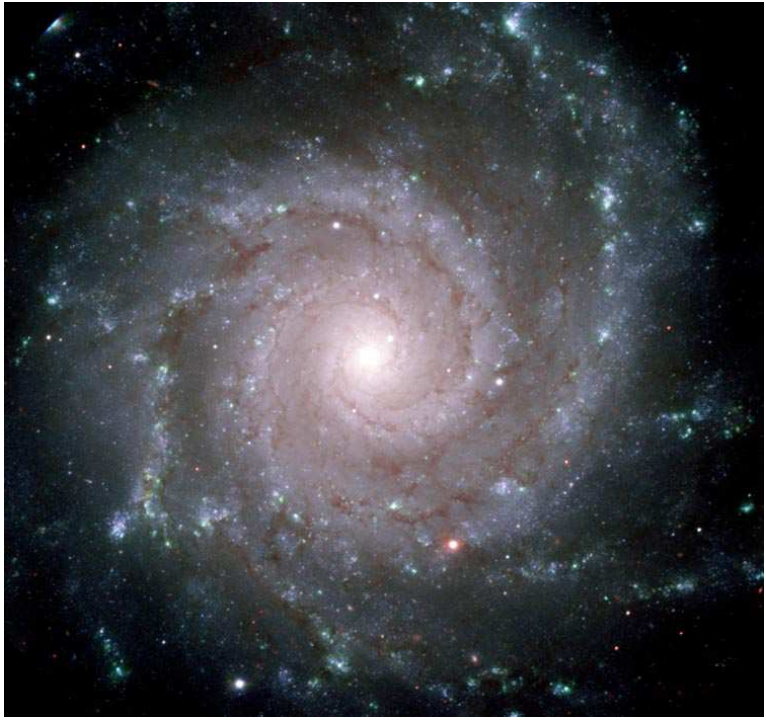}
\caption{Spiral galaxy M74.} \label{M74}
\end{figure}

A galaxy is a collection of $\sim 10^{11}$ stars, which is
gravitationally bound together, and which is $\sim 10^5$ light
years in diameter. Stars within a galaxy are separated by
distances of $\sim 10$ light years, (Rindler 2001, p350). There
are two main types of galaxy: spiral and elliptical. Most of the
stars in a spiral galaxy are concentrated in a flattened disk,
which rotates about its center. The rotation, however, is not
rigid, and the angular velocity of the rotation decreases as a
function of radius. The average rotation period in a spiral galaxy
is about $100$ million years, (ibid.). The disk of a spiral galaxy
is surrounded by a more diffuse `halo' of stars, distributed in a
spherically symmetric fashion. The halo is populated only by
Population $\rm{II}$ stars. Whilst successive generations of stars
form throughout most of the disk in a spiral galaxy, there is
typically a bulge at the center of the disk, and this bulge also
tends to contain Population $\rm{II}$ stars. At the very center of
the disk there is typically a supermassive black hole. A spiral
galaxy possesses an interstellar medium which consists of gas and
dust.\footnote{In relativistic cosmology, `dust' means a
pressure-less fluid. In astronomy and astrophysics, however, dust
means tiny grains of solid matter (Nicolson 1999, p157).}

The stars in the disk of a spiral galaxy are not actually
distributed in spiral patterns. Rather, the spirals trace the
brightest regions in the disk of the galaxy. The brightest regions
are those which contain both the O- and B-type stars and the
H$\rm{II}$ regions of ionized hydrogen, created by the ultraviolet
radiation from those high-temperature O and B-type stars. Because
the brightest (high mass) stars are also the stars with the
shortest lifetimes, the bright high-mass stars can only be found
in regions of recent star formation, and these regions happen to
be the spiral arms.

In an elliptical galaxy, the stars are uniformly distributed
within a $3$-dimensional ellipsoid, and there is no overall
rotation. In fact, the velocities of the stars in an elliptical
galaxy are randomly distributed in all directions. Much like the
halo of a spiral galaxy, there appears to be only one generation
of star formation in an elliptical galaxy.

Our own galaxy, the Milky Way, is a so-called barred-spiral, in
which the spiral arms emanate from the ends of an apparently rigid
bar of luminous matter, (Nicolson 1999, p206). There appear to be
two main spiral arms in the Milky Way: the Norma arm and the
Sagittarius arm. In addition, there are a number of smaller arms,
and the Sun can be found within the Orion arm or `spur', which
lies outside the Sagittarius arm, and inside the Perseus arm
(Nicolson p202). The Sun resides inside a `bubble' three-hundred
light years in diameter, in which the interstellar medium has been
cleared by a supernova explosion (Smolin 1997, p124). From this
perspective, of the $10^{11}$ stars in our galaxy, only about
$7000$ can be seen with the naked eye, (Rindler p350). Moreover,
only $1\%$ of the light which falls upon the Earth comes from
beyond our galaxy, (Disney 2000, p4).

The mutual gravitation of the stars in a galaxy define an escape
velocity, and, at any one time, most of the stars in a galaxy will
be moving with a velocity less than the escape velocity. In this
sense alone, a galaxy cannot be treated as a collection of
isolated, independent entities. Rather, a galaxy must be
represented as a system, in the sense that it constitutes a
collection of mutually interacting parts. In a spiral galaxy in
particular, one can treat the stars as the members of a
population, and the interstellar medium as the environment with
which that population interacts, via feedback loops. In this
sense, it has been suggested that a spiral galaxy can be treated
as an ecological system, (Smolin 1997, Chapter 9). Perhaps,
however, it would be more accurate to say that biological
ecosystems and spiral galaxies are both instances of the same type
of formal system. In other words, there is a type of formal system
which contains a population, an environment, and a set of feedback
loops between the population and the environment, and spiral
galaxies provide instances of this system-type just as much as
biological ecosystems do.

Smolin asserts that ``There are processes by which the matter of
the interstellar medium is converted into stars and there are
processes by which matter is returned from the stars to the
interstellar medium. To understand what a galaxy is, and
especially to understand it as a system, is then primarily to
understand the processes that govern the flow of matter and energy
between the stars and the interstellar medium," (1997 p118). He
points out (p124) that the interstellar medium is a system far
from thermodynamic equilibrium, consisting of a number of
different components of different densities, temperatures, and
compositions. The different components include normal atomic gas;
cold, dense Giant Molecular Clouds (GMCs); and regions of hot,
dilute plasma, otherwise known as H$\rm{II}$ regions, denoting
ionized hydrogen. The relative amount of matter in these different
components remains approximately constant over time. Smolin
asserts that for a system to be in such a stable, but
far-from-equilibrium state, there must be processes which cycle
the material among the different components, and the rates of
these processes must be controlled by feedback mechanisms.

The arms of a spiral galaxy trail behind with respect to the
direction of galactic rotation, hence the obvious explanation for
their shape is simply the fact that the angular rotation velocity
decreases with radial distance from the centre of a galaxy.
However, at the very least, this \emph{differential rotation}
cannot be the sole explanation because spiral galaxies have
existed for at least $10$ billion years, and given a typical
rotation period of $100$ million years, this means that a typical
spiral galaxy has completed at least $100$ rotations, and ``would
long ago have wrapped its arms to extinction if they had been
created early in its history and had not been sustained by some
ongoing process," (Nicolson 1999, p208). Observations also suggest
that the spiral arms do not rotate with the galaxy, but at a
slightly slower rate. Note carefully, however, that an explanation
which uses differential rotation in part is not excluded; it may
still be possible to explain the spiral arms by invoking
differential rotation and an ongoing process which continually
creates the arms. What cannot be done is to explain spiral arms
simply in terms of differential rotation acting upon the initial
conditions under which galaxies are created. The burden of
explanation has been shifted from the initial conditions under
which a spiral galaxy is created to the ongoing processes which
continually operate within such a galaxy.

\begin{figure}[h]
\centering
\includegraphics[scale = 0.4]{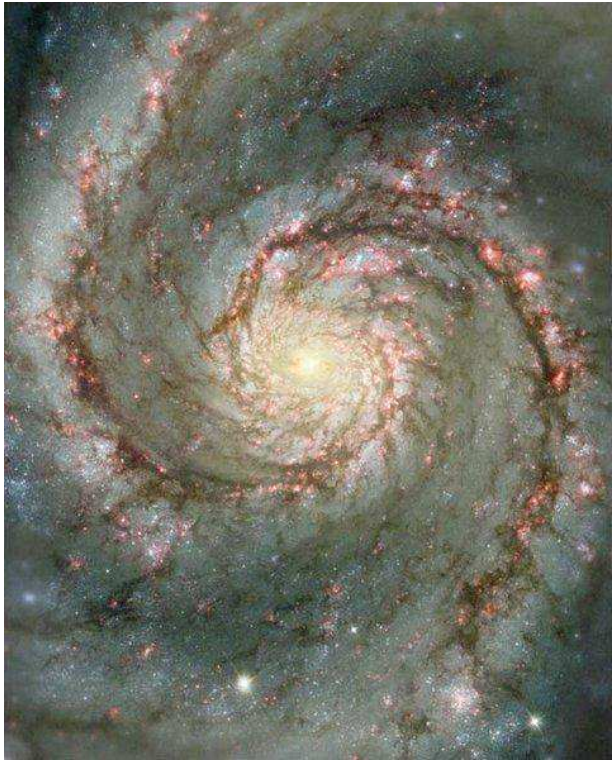}
\caption{Hubble Space Telescope image of M51, the Whirlpool
galaxy.} \label{M51}
\end{figure}

\begin{figure}[h]
\centering
\includegraphics[scale = 0.4]{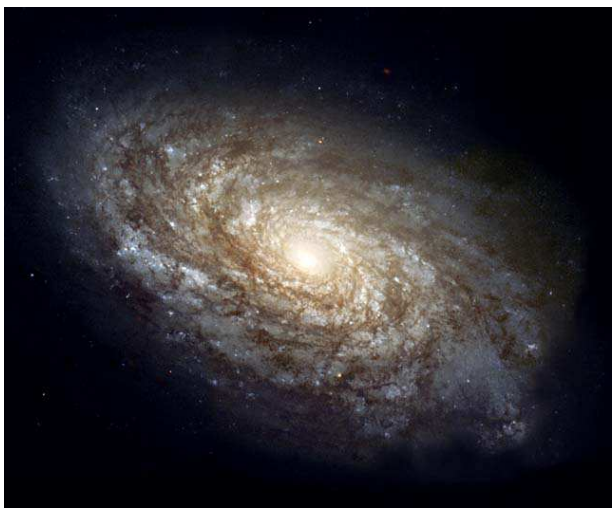}
\caption{Flocculent spiral galaxy NGC4414.} \label{NGC4414}
\end{figure}

There are two processes which have been suggested to explain the
ongoing formation of spiral arms: density-waves and
self-propagating star formation. As Nicolson (p208) points out,
both processes may participate in spiral arm formation, and may be
of differing importance in different types of spiral galaxy. There
are so-called `grand-design' spirals, which possess thin, long and
well-defined spiral arms (see Figure \ref{M51}); there are
flocculent spirals, which possess many fluffy, poorly-defined
spiral arms (see Figure \ref{NGC4414}); and there is a continuum
of intermediate cases, of which the Milky Way is one such.
Karttunen \textit{et al} point out that ``in multiarmed galaxies
the spiral arms may be short-lived, constantly forming and
disappearing, but extensive, regular, two-armed patterns have to
be more long-lived," (2003 p359). As Nicolson suggests,
density-wave mechanisms are more appropriate for grand-design
spirals, and self-propagating star formation is more appropriate
for flocculent spirals.

It was Bertil Lindblad who suggested in the 1920s that spiral arms
could be produced by density waves propagating in a galaxy. The
peaks and troughs of a density wave are independent of any
particular elements from the medium through which the density wave
is propagating. Thus, the density-wave explanation accepts that
stars pass in and out of the spiral arms. However, it also accepts
that spiral arms are genuine indicators of higher-than-average
stellar density. The postulated density wave travels faster than
the speed of sound in the interstellar medium, and the shock wave
purportedly triggers compression of the interstellar medium, and
the observed star formation in the spiral arms. The spiral arms
therefore trace the brightest regions in the disk of a galaxy
because the brightest regions are those which contain the
shortlived O- and B-type stars. As Tayler states, ``the most
massive stars have such a short lifetime that, by the time they
have ceased to be luminous, the spiral pattern has hardly changed
its position. They should therefore only be found in the spiral
regions\dots Stars of lower mass, such as the Sun, have a main
sequence lifetime which is equal to many rotation periods of the
pattern. Such stars should therefore be observed throughout the
disk with no significant correlation with the present position of
the spiral arms," (1993, p145).

C.C.Lin and Frank Shu developed the density-wave idea further in
1964 by postulating that the stars in a spiral galaxy travel in
slightly elliptical orbits which precess with the passage of
time.\footnote{See Pasha (2004) for a detailed history of these
ideas.} This postulate was complemented by the suggestion of J.
Kalnajs in 1973 that if the orientation of the major axis of these
ellipses varies by a small angular increment at increasing
distance from the galactic centre, then the ellipses fail to be
uniformly separated, and where they come close together, they
produce the appearance of grand-design spiral arms.

Kalnajs's explanation of spiral arms can be seen as a
\emph{structural explanation}, in the sense that the spiral arms
are represented as structural elements in a geometrical model, and
these structural elements are linked to other structural elements
in the model, namely the precessing elliptical orbits, each
aligned at a small angle to the one inside it.

Nicolson suggests that density-wave spiral arms ``are probably
sustained either by the asymmetric gravitational field associated
with a central bar structure (typical of grand-design spirals) or
by gravitational disturbances caused by neighboring galaxies, or
by a combination of both," (1999, p208).

In the 1970s Mueller and Arnett suggested an alternative
explanation for spiral arms, which rejects the idea that all
spiral arms are density waves, and postulates instead that at
least some are self-propagating waves of star formation. Unlike
density waves, these waves purportedly do not require one to
postulate external mechanisms, such as the asymmetric
gravitational field of the central bar, or the tidal forces caused
by other galaxies. Let us, then, examine star formation processes
in a little more detail.

As stated in Section 1, the mass and chemical composition of stars
vary, and these are the characteristics which determine the
history and lifetime of a star. New stars are continually born in
a spiral galaxy, but they are not born from other stars, and
consequently the mass of a new star is not in any sense inherited.
The first generation of stars must have formed from clouds of
almost pure hydrogen and helium. Subsequent generations of stars
are observed to form from the cold, dense Giant Molecular Clouds
of the interstellar medium. Given an environment which already
contains surrounding stars, GMCs are only able to remain cool
because they contain dust, which acts as a shield to starlight,
and because they contain organic molecules, which are able to
radiate excess heat (Smolin p110). Organic molecules contain
carbon, a by-product of the nuclear fusion processes in previous
generations of stars.

The star-formation processes within GMCs are thought to be
triggered by the shock-waves from the supernovae explosions of
nearby high-mass stars, formed within neighbouring GMCs $10$
million years previously, or less. The shock-waves purportedly
compress the interstellar medium, and instigate star formation.
Because one spell of star-formation is triggered by the death of
high-mass stars from a previous spell of star-formation, this
process is referred to as self-propagating star-formation (Smolin
p128). Indeed, the inner lanes of spiral arms are often observed
to be dark and dusty, while the outer sides contain the
star-forming regions. This suggests the star formation is directed
against the direction of rotation, which would explain why spiral
arms rotate slightly slower than a host galaxy.

Nicolson (p208) suggests that after a burst of such
self-propagating star formation, differential rotation stretches
the region into a spiral arm. He suggests that in a flocculent
spiral, random bursts of star formation produce numerous spiral
arms by this mechanism, each of which fades away after star
formation has ceased. Subsequent star formation elsewhere in the
galaxy then creates more spiral arms. This differs somewhat from
Smolin's notion of self-propagating star formation, in which he
states that ``the waves of star formation neither die out, nor
grow uncontrollably, but propagate at exactly the right rate to
persist in the galaxy indefinitely," (1997 p135).

The debate between the two different theories which attempt to
explain spiral structure is particularly interesting when one
appreciates that they take opposite stances on the basic cause and
effect relationships which operate in a spiral galaxy. The density
wave theory represents a spiral to be a spiral of density waves,
and explains star formation as the consequence of the density
waves. In contrast, the theory of self-propagating star formation
represents spiral structure to be the consequence of waves of star
formation, (Tayler 1993, p145).

It would not be correct to say that the chemical composition of a
new star is independent of the chemical composition of previous
generations of stars. The chemical composition of the interstellar
medium in a spiral galaxy is changing with the passage of time.
The material expelled from one generation of stars provides a
high-metallicity contribution to the medium from which the next
generation of stars is composed. High metallicity purportedly
inhibits the formation of higher-mass stars, hence the relative
birthrate of lower-mass stars in a spiral galaxy increases with
the passage of time. Accordingly, the birthrate of stars in the
solar neighbourhood is claimed to be inversely proportional to
mass $m$ according to the Salpeter form, $m^{-7/3}$, (Tayler 1993,
p149). It is also suggested that the formation rate of high-mass
stars is less than the formation rate of low-mass stars because
the energy from a newly-created high-mass star heats up the medium
from which it was born to such an extent that the star formation
process halts, (Smolin p127). In general statistical terms, the
number density $N_m$ of type-$m$ objects in such a population is
the product $N_m = B_m \cdot l_m$ of the type-$m$ birthrate $B_m$
with the type-$m$ life-time $l_m$. If we let $N_m$ denote the
number density of mass-$m$ stars, then it follows that because
small mass stars have a greater lifetime, even if they have the
same birth-rate as high-mass stars, they will come to dominate the
population. The eventual domination of lower-mass stars is
therefore the consequence of both statistics and the physics of
star formation processes. In a spiral galaxy, however, where the
birthrates are far from being spatially homogeneous, and the
majority of star formation occurs in the spiral arms, one cannot
infer birthrates from the known lifetimes of stars and the
observed number densities in regions which are remote from the
star formation in spiral arms.

The population of stars in a spiral galaxy is therefore a type of
population in which: (i) there are variable characteristics
distributed within the population, and a small subset of these
characteristics define different types within the population; (ii)
each member of the population has a finite lifetime, determined by
the values of the type-defining characteristics; and (iii) new
members of the population are not reproduced from existing members
of the population, i.e., new members of the population are born
without having any parent(s) in the existing population. Suppose
in addition that (iv) population members of each type are created
at approximately the same rate, or short-lifetime members are
created at a lower rate than long-lifetime members. Such a
population evolves to be dominated by the long-lifetime objects.
Such a population is neither evolving randomly, nor is it evolving
by natural selection.


\begin{thebibliography}{99}
\bibitem{CoxSalp}
Cox, J.P., Salpeter, E.E. (1964). Equilibrium Models for
Helium-Burning Stars. III. Semi-Degenerate Stars of Small Mass,
\emph{Astrophysical Journal}, vol. 140, p.485.
\bibitem{Disney00}
Disney, M.J. (2000). \emph{The case against cosmology},
arXiv:astro-ph/0009020 v1 1 Sep 2000.
\bibitem{GabGrot}
Gabriel, M., Noels-Gr\"otsch, A. (1968). Stabilité séculaire des
étoiles de carbone pur, \emph{Annales d'Astrophysique}, vol. 31,
p.167.
\bibitem{Hansen78}
Hansen, C.J. (1978). Secular stability: applications to stellar
structure and evolution. \emph{Annual review of astronomy and
astrophysics}, volume 16, pp15-32.
\bibitem{Hansen04}
Hansen, C.J., Kawaler, S.D., Trimble, V. (2004). \emph{Stellar
Interiors: Physical Principles, Structure and Evolution}, 2nd
Edition. Berlin-Heidelberg-New York: Springer Verlag.
\bibitem{Kah78}
K\"ahler, H. (1978). The Vogt-Russell theorem, and new results on
an old problem, pp303-311 in \emph{The HR diagram - The 100th
anniversary of Henry Norris Russell}, A.G. Davis Philip and
D.S.Hayes (eds.). Dordrecht: Reidel.
\bibitem{Karttunen03}
Karttunnen, H., Kr\"oger, P., Oja, H., Poutanen, M., Donner, K.J.
(2003). \emph{Fundamental astronomy}, Fourth Edition.
Berlin-Heidelberg-New York: Springer Verlag.
\bibitem{KippWei90}
Kippenhahn, R., Weigert, A. (1990). \emph{Stellar Structure and
Evolution}. Berlin-Heidelberg-New York: Springer Verlag.
\bibitem{Nicolson99}
Nicolson, I. (1999). \emph{Unfolding our universe}. Cambridge:
Cambridge University Press.
\bibitem{Pasha04}
Pasha, I.I. (2004). Density-wave spiral theories in the 1940s, I.
arXiv:astro-ph/0406142
\bibitem{Rindler01}
Rindler, W. (2001). \emph{Relativity: Special, General, and
Cosmological}. Oxford: Oxford University Press.
\bibitem{Russell}
Russell, H.N., Dugan, R.S., Stewart, J.Q. (1927). \emph{Astronomy}
Vol. 2. Boston: Ginn and Co.
\bibitem{Smolin97}
Smolin, L. (1997a). \emph{The Life of the Cosmos}. London:
Weidenfeld and Nicolson.
\bibitem{Tayler93}
Tayler, R.J. (1993). \emph{Galaxies: structure and evolution}.
Cambridge: Cambridge University Press.
\bibitem{Tayler94}
Tayler, R.J. (1994). \emph{The Stars: their structure and
evolution}. Cambridge: Cambridge University Press.
\bibitem{Vogt}
Vogt, H. (1926). Die Beziehung zwischen den Massen und den
absoluten Leuchtkr\"aften der Sterne, \emph{Astronomische
Nachrichten}, volume 226, p.301.
\end{thebibliography}
\end{document}